\documentclass[a4paper,USenglish,cleveref, autoref]{lipics-v2019}

\title{Formalizing the Solution to the Cap Set Problem}

\author{Sander R. Dahmen}{Department of Mathematics, Vrije Universiteit Amsterdam, The Netherlands}{s.r.dahmen@vu.nl}
{https://orcid.org/0000-0002-0014-0789}
{NWO Vidi grant No. 639.032.613, New Diophantine Directions.}

\author{Johannes H\"olzl}{Department of Computer Science, Vrije Universiteit Amsterdam, The Netherlands}{johannes.hoelzl@posteo.de}
{https://orcid.org/0000-0003-0869-9250}
{ERC grant agreement No. 713999, Matryoshka.}

\author{Robert Y. Lewis}{Department of Computer Science, Vrije Universiteit Amsterdam, The Netherlands}{r.y.lewis@vu.nl}
{https://orcid.org/0000-0002-5266-1121}
{ERC grant agreement No. 713999, Matryoshka.}

\authorrunning{S.\,R.\,Dahmen, J.\,H\"olzl, and R.\,Y.\,Lewis}

\Copyright{Sander R. Dahmen, Johannes H\"olzl, and Robert Y. Lewis}

\ccsdesc[500]{Theory of computation~Logic and verification}
\ccsdesc[500]{Theory of computation~Type theory}
\ccsdesc[300]{Mathematics of computing~Number-theoretic computations}
\ccsdesc[300]{Computing methodologies~Combinatorial algorithms}
\ccsdesc[300]{Software and its engineering~Formal methods}

\keywords{formal proof, combinatorics, cap set problem, Lean}

\category{}

\relatedversion{}

\supplement{Links to our formalization and supporting documents are hosted at the URL \url{https://lean-forward.github.io/e-g/}}

\funding{}

\acknowledgements{We are grateful to the Lean \mathlib\ maintainers and contributors on whose work this project is based. We thank Jeremy Avigad and Jasmin Blanchette for helpful comments on this paper, Dion Gijswijt for suggesting an improvement to our asymptotic bound argument, and Manuel Eberl for pointing us to related work.}

\nolinenumbers 

\hideLIPIcs  

\EventEditors{John Harrison, John O'Leary, and Andrew Tolmach}
\EventNoEds{3}
\EventLongTitle{10th International Conference on Interactive Theorem Proving (ITP 2019)}
\EventShortTitle{ITP 2019}
\EventAcronym{ITP}
\EventYear{2019}
\EventDate{September 9--12, 2019}
\EventLocation{Portland, OR, USA}
\EventLogo{}
\SeriesVolume{141}
\ArticleNo{21}

\usepackage{scalefnt}
\usepackage{ucs}
\inputencoding{utf8x}
\usepackage{upgreek}
\usepackage{courier} 
\usepackage{tikz}
\usepackage{setdeck}

\usetikzlibrary{arrows,positioning,calc,shapes.geometric,patterns}

\definecolor{keywordcolor}{rgb}{0.7, 0.1, 0.1}   
\definecolor{tacticcolor}{rgb}{0.1, 0.2, 0.6}    
\definecolor{commentcolor}{rgb}{0.4, 0.4, 0.4}   
\definecolor{symbolcolor}{rgb}{0.0, 0.1, 0.6}    
\definecolor{sortcolor}{rgb}{0.1, 0.5, 0.1}      

\newcommand{\lean}[1]{\lstinline[language=lean]{#1}}

\newcommand{\RR}{\mathbb{R}}
\newcommand{\QQ}{\mathbb{Q}}
\newcommand{\ZZ}{\mathbb{Z}}
\newcommand{\NN}{\mathbb{N}}
\newcommand{\CC}{\mathbb{C}}
\newcommand{\Fq}{\mathbb{F}_q}
\newcommand{\mathlib}{{\tt mathlib}}

\colorlet{White}{white}
\colorlet{OliveGreen}{black!30!green}
\colorlet{PineGreen}{black!30!green}
\colorlet{RoyalPurple}{violet}
\colorlet{Red}{red}

\begin{document}

\maketitle

\begin{abstract}
In 2016, Ellenberg and Gijswijt established a new upper bound on the size of subsets of $\mathbb{F}^n_q$ with no three-term arithmetic progression. This problem has received much mathematical attention, particularly in the case $q = 3$, where it is commonly known as the \emph{cap set problem}. Ellenberg and Gijswijt's proof was published in the \emph{Annals of Mathematics} and is noteworthy for its clever use of elementary methods. This paper describes a formalization of this proof in the Lean proof assistant, including both the general result in $\mathbb{F}^n_q$ and concrete values for the case $q = 3$. We faithfully follow the pen and paper argument to construct the bound. Our work shows that (some) modern mathematics is within the range of proof assistants.
\end{abstract}

\vspace{.2cm}
[This is a preprint of a paper that will appear in the proceedings of ITP 2019.]

\section{Introduction}
\label{section:introduction}

As proof assistants improve and their libraries grow, these tools are increasingly used to formalize results at the cutting edge of computer science. At some prestigious conferences such as \emph{Principles of Programming Languages} (POPL), it is common for papers establishing new metatheoretical results about programming languages to be accompanied by formal proofs. In the field of mathematics, however, the picture looks very different. Even though early proof assistants were developed by and for mathematicians~\cite{deBruijn:83, matuszewski:05}, there are still very few mathematicians who use these tools in their work. With a small number of noteworthy exceptions (e.g.\ Gou\"ezel and Schur \cite{gouezel:18} and Hales, et al.\ \cite{hales:17}), no current work in pure mathematics work gets formalized; most of the results formalized in papers at \emph{Interactive Theorem Proving} (ITP) or \emph{Certified Programs and Proofs} (CPP) have already made it into undergraduate or introductory graduate textbooks.

Researchers often point to the \emph{depth} of mathematical theory to explain this difference. While programming language formalizations can be sprawling and difficult, they rarely depend on large background libraries, and often involve repetitive arguments that are amenable to automation. In comparison, mathematics builds upwards on centuries of earlier work, and one cannot formalize modern results without first formalizing the necessary foundation. The few existing formal developments of cutting-edge mathematics tend to focus on results that are difficult to verify by hand---justifying the effort needed to develop libraries---or fall in subfields of mathematics where the background theory is less intimidating.

The combinatorial proof described in this paper belongs in the latter category. Let $G$ be an abelian group. A three-term \emph{arithmetic progression} of elements of $G$ is a sequence $a, a + g, a + g + g$ where $a,g\in G$ and $g$ is nonzero. Let $r_3(G)$ denote the cardinality of a largest subset of $G$ containing no three-term arithmetic progression. We will focus on the group $(\ZZ/3\ZZ)^n = \{(a_1,\ldots,a_n) \mid a_i \in \{0,1,2\}\}$, where vector addition is pointwise and modulo 3; a subset of this group with no three-term arithmetic progression is known as a \emph{cap set}. The \emph{cap set problem} asks whether there is a constant $c < 3$ such that $r_3((\ZZ/3\ZZ)^n)$ grows in $n$ no faster than $c^n$.

\begin{figure}
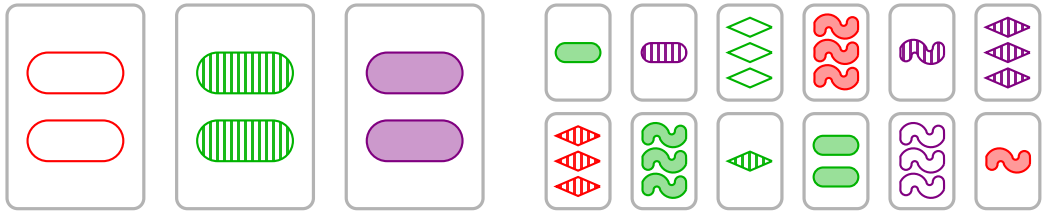

\centering
\begin{subfigure}[t]{.48\columnwidth}
\centering
\setlength{\tabcolsep}{4pt}
\begin{tabular}{ccc}
\setcard{2}{1}{1}{2}{.45} & \setcard{2}{2}{2}{2}{.45} & \setcard{2}{3}{3}{2}{.45}
\end{tabular}

\subcaption{A valid triple. Each card has the same shape and the same number of shapes. Each card has a different color and a different fill.}
\end{subfigure}
\hfill
\begin{subfigure}[t]{.49\columnwidth}
\centering
\setlength{\tabcolsep}{2pt}
\begin{tabular}{cccccc}

\setcard{1}{3}{2}{2}{.21} & \setcard{1}{2}{3}{2}{.21} & \setcard{3}{1}{2}{1}{.21} &
\setcard{3}{3}{1}{3}{.21} & \setcard{1}{2}{3}{3}{.21} & \setcard{3}{2}{3}{1}{.21} \\

\setcard{3}{2}{1}{1}{.21} & \setcard{3}{3}{2}{3}{.21} & \setcard{1}{2}{2}{1}{.21} &
\setcard{2}{3}{2}{2}{.21} & \setcard{3}{1}{3}{3}{.21} & \setcard{1}{3}{1}{3}{.21}
\end{tabular}
\setlength{\tabcolsep}{6pt}
\subcaption{A collection of twelve cards that contains no valid triple.}
\end{subfigure}

\caption{The cap set problem can be interpreted in the game Set, where it concerns an upper bound on the size of a collection of cards that contains no valid triple.}

\label{figure:set}

\end{figure}

Readers familiar with the card game Set (Figure~\ref{figure:set}) may understand the cap set problem in different terms. A card in Set has four features, where each feature has three possible values. (A card has one, two, or three copies of a shape; the shape is an oval, a diamond, or a squiggle; the shape is solid, striped, or empty; the shape is purple, red, or green.) A triple of cards is said to be \emph{valid} if, for each feature, either all three cards have the same value or all three cards have different values. During game play, players search a collection of cards for valid triples. The number $r_3((\ZZ/3\ZZ)^4)$ is the maximum size of a collection of distinct cards in which no valid triples can be found, and the cap set problem concerns the growth rate of this value as the number of features is increased.

The cap set problem is surprisingly difficult to analyze and has attracted attention over the past decades from leading combinatorialists. Croot, Lev, and Pach~\cite{croot:17} solved a closely related problem in 2016. Building on their work, Ellenberg and Gijswijt soon showed that $r_3((\ZZ/3\ZZ)^n)$ is $o(2.756^n)$, a major breakthrough. In fact, they proved a more general result about finite fields. Their 2017 paper in the \emph{Annals of Mathematics}~\cite{ellenberg:gijswijt:17} is noteworthy in that the core of the proof does not use any complicated theoretical machinery. Rather, it relies on a clever shift of context, casting the problem in terms of polynomials of bounded degree. While their final proof of the asymptotics does make use of relatively high-powered methods, Tao~\cite{tao:16} and Zeilberger~\cite{zeilberger:16} indicate how these calculations can be made elementary. We also note that Tao~\cite{tao:16} reformulates Ellenberg and Gijswijt's proof in a more symmetric way, using what is now called ``slice rank.'' Although this is arguably a more natural way to express things, the underlying arguments are essentially the same.

This paper describes a formalization of Ellenberg and Gijswijt's argument, carried out in the Lean proof assistant. While unavoidably more verbose, our computation of an upper bound for $r_3((\ZZ/p\ZZ)^n)$ faithfully follows Ellenberg and Gijswijt's proof. To verify the asymptotics, we work out a new elementary argument (inspired by Zeilberger's approach and a suggestion by Gijswijt). Ellenberg and Gijswijt use a technique known as the \emph{polynomial method} to translate the problem to one about vector spaces of polynomials. We expect that our library contributions will be useful for proving other results that follow this approach.

A recent project begun at the Vrije Universiteit Amsterdam aims to bring together traditional mathematicians, formalizers, and tool developers to incorporate modern number theory into proof assistants.\footnote{\url{https://lean-forward.github.io/}} The current paper shows that the goals of this project are within reach: we have formalized a paper published in the \emph{Annals} less than two years ago.

The more general components of our formalization have been incorporated into the Lean mathematics library \mathlib, which is available on GitHub.\footnote{\url{https://github.com/leanprover-community/mathlib/}} The remainder of the formalization can be found with the supplementary material linked at the beginning of this paper. The code blocks presented below should be read as schematic, not literal. We sometimes change names, remove namespaces, omit universe levels, and swap implicit and explicit arguments for the sake of formatting and presentation.


\section{Mathematical Background}
\label{section:background}

Ellenberg and Gijswijt study a generalization of the cap set problem that holds for arbitrary finite fields  (including $\ZZ/p\ZZ$ for any prime $p$). For the rest of this discussion, we fix a positive integer $n$ and prime power $q$, and let $\Fq$ denote a finite field with cardinality $q$.

For $d \in \RR$ with $0 \leq d \leq (q-1)n$, consider all $n$-variable monomials whose degree in each variable is at most $q-1$ and whose total degree is at most $d$, i.e.
\[
M_n^d : =\left\{\prod_{i=1}^n x_i^{a_i} \in \Fq[x_1,\ldots,x_n] \enspace \middle\vert \enspace 0 \leq a_i \leq q-1 \text{ and }  \sum_{i=1}^n a_i \leq d\right\}.
\]
Let $m_d:=|M_n^d|$. Ellenberg and Gijswijt~\cite[Theorem 4] {ellenberg:gijswijt:17} establish an upper bound for the size of generalized cap sets in terms of $m_{(q-1)n/3}$.

\begin{theorem}[Ellenberg--Gijswijt]
\label{theorem:bound}
Let $\alpha, \beta, \gamma \in \mathbb{F}_q$ such that $\alpha + \beta + \gamma = 0$ and $\gamma \neq 0$. Let $A$ be a subset of $\Fq^n$ such that the equation $\alpha a_1 + \beta a_2 + \gamma a_3 = 0$ has no solutions with $a_1, a_2, a_3 \in A$ apart from those with $a_1 = a_2 = a_3$.
Then $|A| \leq 3m_{(q-1)n/3}$.
\end{theorem}
If $(\alpha, \beta, \gamma)=(1,-2,1)$, then the equation $\alpha a_1 + \beta a_2 + \gamma a_3 = 0$  is equivalent to $a_2-a_1=a_3-a_2$; any solution to this, other than $a_1=a_2=a_3$, corresponds to a three term arithmetic progression.

To answer the cap set problem, it remains to determine good asymptotics for $m_{(q-1)n/3}$ as $n$ tends to $\infty$.

\begin{theorem}\label{theorem:asymptotics}
For every $q$ there exists $c \in \RR$ with $0 < c < q$ such that $m_{(q-1)n/3} = \mathcal{O}(c^n)$ as $n \to \infty$.
\end{theorem}
Thus, with notation from Theorem~\ref{theorem:bound}, $|A|= \mathcal{O}(c^n)$ for some $0<c<q$.
For particular values of $q$ we can write down explicit values of $c$. In the case of the original cap set problem, where $q=3$ (and $\alpha=\beta=\gamma=1$, also noting that $-2=1$ in $\ZZ/3\ZZ$), the proof method yields the following theorem; the exact value $c$ already appears in Zeilberger~\cite{zeilberger:16}.

\begin{theorem}\label{theorem:cap set}
Let $c:=\frac{3}{8} \sqrt[3]{207+33\sqrt{33}} < 2.755105$. Then $r_3\left((\ZZ/3\ZZ)^n\right) \leq 3 c^n$, and thus $r_3\left((\ZZ/3\ZZ)^n\right)= o( 2.755105^n)$ (as $n \to \infty$).
\end{theorem}

The proof of Theorem~\ref{theorem:bound} follows the \emph{polynomial method}. (For a general introduction to the polynomial method, see e.g.\ Guth~\cite{guth:16} or Tao~\cite{tao:14}.) Broadly speaking, this approach aims to analyze finite combinatorial objects by describing them through a system or space of polynomials. Techniques from algebraic geometry, or sometimes algebraic topology or simply linear algebra, can then be employed to study these polynomials; the results should translate back to properties of the original combinatorial objects of interest.

The polynomial method has been employed over the last decade to solve a large variety of open problems in arithmetic combinatorics and number theory. However, the scope and limitations of the method are still not well understood. In particular, its applicability to the cap set problem was unexpected, at least until the breakthrough of Croot, Lev, and Pach~\cite{croot:17}. The main approach to the cap set problem for the previous half century was through Fourier theory methods.

We sketch here an overview of the proof of Theorem~\ref{theorem:bound}; more details can be found in Section~\ref{section:bound}.
Let $\alpha, \beta, \gamma$, and $A$ be as stated in the theorem.
We introduce the $\Fq$-vector space spanned by $M_n^d$, i.e.
\[
S_n^d:=\left\{\sum_{m \in M_n^d} c_m m \enspace\middle\vert\enspace c_m \in \Fq\right\}.
\]
Consider the $\Fq$-vector subspace $V$ of $S_n^d$ consisting of all polynomials $p \in S_n^d$ that vanish on the complement of $-\gamma A=\{-\gamma a \mid a \in A\}$ inside $\Fq^n$, i.e.
\[
V:= \{p \in S_n^d \mid \forall a \in \Fq^n \setminus (-\gamma A),\ p(a)=0\}.
\]
This is the setup of the polynomial method, the idea being that this space of polynomials $V$ contains valuable information on $|-\gamma A|=|A|$ via $\dim(V)$. The strategy is to get good lower and upper bounds on $\dim(V)$. Namely, it holds that
\begin{equation}\label{eqn:bounds dim V}
\dim(V) \geq m_d-q^n+|A| \qquad \text{and} \qquad \dim(V) \leq 2m_{d/2}.
\end{equation}
The lower bound is reasonably straightforward: it follows from rank-nullity and the remark that $|\Fq^n\setminus(-\gamma A)|=q^n-|A|$. The upper bound is more involved; the key to it is the following.
\begin{proposition}[Proposition 2 from~\cite{ellenberg:gijswijt:17}]\label{prop 2 E-G}
Let $A \subseteq \Fq^n$ and $\alpha, \beta, \gamma \in \Fq$ with $\alpha + \beta + \gamma=0$.
Let $P \in S_n^d$ such that for all $a,b \in A$ with $a\not=b$ we have $P(\alpha a+ \beta b)=0$.
Then
\[|\{a \in A \mid P(-\gamma a)\not=0\}|\leq 2m_{d/2}.\]
\end{proposition}
In addition, an elementary combinatorial argument gives us
\begin{equation}\label{eqn:combinatorial bound}
q^n-m_d \leq m_{(q-1)n-d}.
\end{equation}
Combining~\eqref{eqn:bounds dim V} and~\eqref{eqn:combinatorial bound} and taking $d=2(q-1)n/3$ gives us Theorem~\ref{theorem:bound}, i.e.
\[|A| \leq 3m_{(q-1)n/3}.\]

To establish the asymptotic behavior of this bound, Ellenberg and Gijswijt apply Cram\'er's theorem on large deviations. Tao~\cite{tao:16} describes a more elementary approach via Stirling's approximation for the factorial function. Zeilberger~\cite{zeilberger:16} gives another even more elementary approach using recurrence sequences. Inspired by Zeilberger's paper, we worked out yet another approach, which lends itself very well to formalization in Lean. This was the initial approach we followed through; it is briefly described in Appendix~\ref{appendix}. Finally, thanks to a remark from Dion Gijswijt on our preprint, we arrive at a further significant simplification of the asymptotics proof, which we present below.

Our starting point is the combinatorial observation
\begin{equation}\label{eqn:comb id}
m_d=\sum_{j=0}^{\lfloor d \rfloor} c_j^{(n)}
\end{equation}
where $c_j^{(n)}$ is the coefficient of $x^j$ in the polynomial $\left(1+x+\ldots x^{q-1}\right)^n$.
Let $r \in \RR$ with $0<r<1$ and write $e:=\lfloor (q-1)n/3 \rfloor$. Note that the $c_j^{(n)}$ are nonnegative and that $r^e \leq r^j$ for integers $0\leq j \leq e$.
Now
\[
m_{(q-1)n/3} \cdot r^e
= \sum_{j=0}^e c_j^{(n)} r^e
\leq  \sum_{j=0}^e c_j^{(n)} r^j
\leq  \sum_{j=0}^{(q-1)n} c_j^{(n)} r^j
= \left(1+r+\ldots+r^{q-1} \right)^n.
\]
Dividing by $r^e\geq (r^{(q-1)/3})^n$ and defining
\begin{equation}\label{eqn:def Crq}
C_{r,q}:=\frac{1+r+\ldots + r^{q-1}}{r^{(q-1)/3}}
= \frac{1-r^q}{(1-r) r^{(q-1)/3}}
\end{equation}
we arrive at our main asymptotics estimate
\[m_{(q-1)n/3} \leq C_{r,q}^n.\]
Elementary analysis gives us that for every $q>1$ there exists some $0<r<1$ such that $C_{r,q}<q$, yielding Theorem~\ref{theorem:asymptotics}. Specializing at $q=3$ and $r=(\sqrt{33}-1)/8$ gives the precise version of the cap set problem in Theorem~\ref{theorem:cap set}.
Similarly, minimizing $C_{r,q}$ for other values of $q$ immediately leads to other growth rates, including those given by Zeilberger~\cite{zeilberger:16}.

\section{Lean and its Mathematics Library}
\label{section:lean}


The Lean proof assistant, developed principally by Leonardo de Moura, was first released in 2014~\cite{demoura:14}. Lean implements a version of the calculus of inductive constructions (CIC)~\cite{coquand:90} with support for quotient types and classical reasoning. Since the release of Lean 3 in 2017~\cite{demoura:17}, there has been a concerted effort to develop \mathlib, a comprehensive library for use in mathematics and computer science~\cite{carneiro:18}. This library is built on the latest release of Lean, version 3.4.2. Some of the text in this section is adapted from a paper by the third author~\cite{lewis:19}, which describes another formalization based on \mathlib.

The datatypes available in \mathlib\ include the concrete types commonly found in mathematics, among them $\NN$, $\ZZ$, $\QQ$, $\RR$, and $\CC$; finite sets and multisets over a base type; univariate and multivariate polynomials; and embeddings and isomorphisms between types. The algebraic hierarchy of mathlib is designed using \emph{type classes}, which endow a base type with extra structure in the forms of operations, properties, and notation~\cite{spitters:11, wadler:89}. Lean's type class resolution mechanism automatically manages inheritance between type classes (Figure~\ref{figure:typeclass}). If a type class \lstinline{T'} extends (directly or by transitivity) a type class \lean{T}, any theorem proved over \lean{T} will apply to any type that instantiates \lean{T'}. The algebraic hierarchy begins with semigroups and monoids and extends to rich structures including fields, Noetherian rings, and principal ideal domains. Van Doorn, von Raumer, and Buchholz~\cite{vandoorn:17} also explain how type classes are used to define an algebraic hierarchy in Lean.

\begin{figure}
\begin{lstlisting}[backgroundcolor=\color{white}]
class semigroup (α : Type) extends has_mul α :=
(mul_assoc : ∀ a b c : α, a * b * c = a * (b * c))

class monoid (α : Type) extends semigroup α, has_one α :=
(one_mul : ∀ a : α, 1 * a = a) (mul_one : ∀ a : α, a * 1 = a)

class group (α : Type) extends monoid α, has_inv α :=
(mul_left_inv : ∀ a : α, a⁻¹ * a = 1)

lemma one_inv (α : Type) [group α] : 1⁻¹ = (1 : α) :=
inv_eq_of_mul_eq_one (one_mul 1)
\end{lstlisting}

\caption{A sample of the bottom of the algebraic hierarchy. The lemma \lean{one_inv} can be applied to
any\!\lean{\ α} 
for which Lean can infer an instance of \lean{group α}.}
\label{figure:typeclass}
\end{figure}

The project described in this paper makes heavy use of the linear algebra and multivariate polynomial developments in \mathlib. As with the algebraic hierarchy, these developments are built around type classes. The linear algebra theory in particular is modeled after the one found in Isabelle/HOL, reworked to use bundled submodules and bundled linear functions.

The fundamental type class in linear algebra is \lean{module α β}, which assumes a ring structure on \lstinline{α} and an abelian group structure on \lstinline{β}, and endows \lstinline{β} with a well-behaved scalar multiplication operation from \lstinline{α}. When \lstinline{α} is a field, this extends to the type class \lean{vector_space α β}. Many of the typical theorems and constructions from linear algebra are defined over this type class, including the existence of bases, the rank-nullity theorem for linear maps, and the matrix representation of maps between finite-dimensional spaces. General instances establish that a family of vector spaces over an index type forms a vector space itself, and that a field \lstinline{α} instantiates \lean{vector_space α α}; combined, these allow us to consider the type of $n$-tuples of field elements, \lean{fin n → α}, as a vector space over \lstinline{α}.

Polynomials are another important instance of a vector space. Given a type \lstinline{σ} used to index variables, we identify a monomial with a finitely supported function from \lstinline{σ} to \lstinline{ℕ}. A multivariate polynomial is a finitely supported function mapping monomials into a coefficient ring \lstinline{α}. We use the infix notation \lstinline{→₀} for functions of finite support.
\vspace{-.1cm}
\begin{lstlisting}
def mv_polynomial (σ α : Type) [comm_semiring α] := (σ →₀ ℕ) →₀ α
\end{lstlisting}
\vspace{-.1cm}
When \lstinline{α} is a field, this type forms a vector space over \lstinline{α}. Important operations on polynomials include \lean{eval}, which evaluates the polynomial in \lstinline{α} given an assignment \lstinline{σ → α}, and \lean{total_degree}, which computes the maximum degree over all monomials in a polynomial.

Many contributions were made to \mathlib\ in the course of this project. In addition to extending the linear algebra, polynomial, and finitely supported function theories, we added various results about big operators and series, finite sets and multisets, and orders of elements in finite groups (to show, for example, that $a^q = a$ for $a \in \Fq$).

Another type class that plays an important role in our formalization is \lean{fintype α}, which provides functions for listing and counting the elements of \lstinline{α}. The standard finite types instantiate this class, including the type \lean{fin n} of natural numbers less than \lean{n}. When \lstinline{α} and \lstinline{β} instantiate \lean{fintype}, so does the function type \lstinline{α → β}.
%
%


%
%
%
%
%


The \mathlib\ library is designed with a focus on classical logic. Type-valued declarations are defined computably when possible, but classical logic is used freely in propositions. Our formalization is similarly classical.

Readers unused to Lean syntax should note that explicit arguments to declarations are enclosed in parentheses \lean{()}, implicit arguments are enclosed in curly brackets \lean{\{\}}, and type class arguments are enclosed in square brackets \lean{[]}. Only explicit arguments are given by the user when applying a declaration. Implicit arguments are inferred from later arguments and the expected type, and type class arguments are inferred by type class resolution.

Another important feature of Lean syntax is its projection notation. As an example, let terms \lean{F : polynomial α} and  \lean{a : α} be given. The operator
\vspace{-.15cm}
\begin{lstlisting}
polynomial.eval : α → polynomial α → α
\end{lstlisting}
\vspace{-.15cm}
evaluates a polynomial at an argument. Because the head symbol of the type of \lean{F} is \lean{polynomial}, matching the namespace of \lean{eval}, we can abbreviate \lean{polynomial.eval a F} with the more concise \lean{F.eval a}. This notation can be nested:
\vspace{-.15cm}
\begin{lstlisting}
polynomial.eval a (polynomial.derivative F)
\end{lstlisting}
\vspace{-.15cm}
shortens to \lean{F.derivative.eval a}.

\section{The Cap Set Bound}
\label{section:bound}


As described in Section~\ref{section:background}, Ellenberg and Gijswijt's solution to the cap set problem~\cite{ellenberg:gijswijt:17} proceeds in two parts. The first part establishes an upper bound on the size of a cap set in terms of the dimension of a vector space of polynomials; the second part shows the asymptotic behavior of this bound. Our formalization is similarly divided. This section describes the formal construction of the bound, and Section~\ref{section:asymptotics} explains the verification of the asymptotics. Our construction of the bound closely follows Ellenberg and Gijswijt's paper.

At the outset of our efforts, the first author produced a detailed paper proof\footnote{This writeup is available at \url{https://lean-forward.github.io/e-g/}} of the result, drawing from Ellenberg and Gijswijt and from Zeilberger~\cite{zeilberger:16} and adapting the asymptotics part significantly. The most recent approach to this part was added after initially submitting this paper, and was subsequently also formalized. The theorem names in the following sections match the corresponding statements in the paper proof.

The theorems here hold over an arbitrary finite field. We will take a fixed parameter \lstinline{α : Type} instantiating the type classes \lean{[fintype α]} and \lean{[discrete_field α]}, and use \lean{q} to abbreviate the cardinality \lean{fintype.card α}. In this section, we also fix a parameter \lean{n : ℕ}, representing the length of the tuples in the set whose cardinality we will bound.

The goal of this section, then, is to define a function \lean{m} and prove the following theorem, which corresponds to the informal statement of Theorem \ref{theorem:bound} above:

{
\noindent
\begin{minipage}{\linewidth}
\begin{lstlisting}
theorem theorem_12_1 {α : Type} [discrete_field α] [fintype α]
  (n : ℕ) {a b c : α} (hc : c ≠ 0) (habc : a + b + c = 0)
  (hn : n > 0) {A : finset (fin n → α)}
  (ha : ∀ x y z ∈ A, a • x + b • y + c • z = 0 → x = y ∧ x = z) :
  A.card ≤ 3 * m α n (1 / 3 * ((card α - 1) * n))
\end{lstlisting}
\end{minipage}}

Ellenberg and Gijswijt's key insight is to translate the question to one concerning vector spaces of multivariate polynomials. After setting up this translation, this bound will follow from a sequence of intermediate lemmas.

\subsection{Setting Up the Polynomial Method}

The type \lean{mv_polynomial (fin n) α} forms a vector space, by results established in \mathlib\ (Section \ref{section:lean}). We will focus our attention on a particular subspace. We define \lean{M} to be the set of monomials in \lean{n} variables where the exponent of each variable is strictly less than \lean{q}. This set is linearly independent with respect to \lstinline{α}.

\begin{lstlisting}
def M : finset (mv_polynomial (fin n) α) :=
(finset.univ.image
  (λ f : fin n →₀ fin q, f.map_range fin.val rfl)).image
    (λ d : fin n →₀ ℕ, monomial d (1:α))
\end{lstlisting}

For \lean{d : ℚ}, we make the following definitions:
\begin{itemize}
 \item \lean{M'} is the subset of \lean{M} whose elements have total degree at most \lean{d}.
 \item \lean{S'} is the span of \lean{M'}; this is a subspace of \lean{mv_polynomial (fin n) α}.
 \item \lean{m} is the dimension of \lean{S'}.
\end{itemize}
Since \lean{M'} is linearly independent, it follows that the cardinality of \lean{M'} is equal to \lean{m}.

\begin{lstlisting}
def M' (d : ℚ) : finset (mv_polynomial (fin n) α) :=
M.filter (λ m, d ≥ mv_polynomial.total_degree m)

def S' (d : ℚ) : submodule α (mv_polynomial (fin n) α) :=
submodule.span α ((M' d) : set (mv_polynomial (fin n) α))

def m (d : ℚ) : ℕ := (vector_space.dim α (S' d)).to_nat

lemma M'_card (d : ℚ) : (M' d).card = m d
\end{lstlisting}

Much of the following argument will be carried out in a subspace of \lean{S'}. We first describe this subspace generically. Given a subspace of polynomials \lean{T} and a set of vectors \lean{A}, we define \lean{zero_set T A} to be the set of polynomials in \lean{T} that evaluate to 0 at all elements of \lean{A}. By basic properties of polynomial evaluation, this set is a subspace of \lean{T}.

\begin{lstlisting}
parameters (T : subspace α (mv_polynomial (fin n) α))
           (A : finset (fin n → α))

def zero_set : set (mv_polynomial (fin n) α) :=
{p ∈ T.carrier | ∀ a ∈ A, mv_polynomial.eval a p = 0}

def zero_set_subspace : subspace α (mv_polynomial (fin n) α) :=
{ carrier := zero_set,
  zero := ⟨submodule.zero, by simp⟩,
  add := λ _ _ hx hy,
   ⟨submodule.add hx.1 hy.1, λ _ hp, by simp [hx.2 hp, hy.2 hp]⟩,
  smul := λ _ _ hp,
   ⟨submodule.smul hp.1, λ _ hx, by simp [hp.2 hx]⟩ }
\end{lstlisting}

Our target theorem takes as parameters \lean{a b c : α} and \lean{A : finset (fin n → α)} satisfying certain properties, in particular that \lean{c ≠ 0}. Let these terms be given. We define \lean{neg_cA} to be the image of \lean{A} under multiplication by \lean{-c}, and \lean{V} to be the zero set of \lean{S'} with respect to the complement of \lean{neg_cA}.

\begin{lstlisting}
def neg_cA : finset (fin n → α) := A.image (λ z, (-c) • z)

def V : subspace α (S' d) :=
zero_set_subspace (S' d) (finset.univ \ neg_cA)

def V_dim : ℕ := (vector_space.dim α V).to_nat
\end{lstlisting}

Our goal---an upper bound on the cardinality of \lean{A}, in terms of \lean{m}---will follow from a number of lemmas controlling the dimension of \lean{V}.

\subsection{Lemma 1: Bounding the Dimension from Below}

The first lemma establishes a lower bound for the dimension of \lean{V} in terms of \lean{m}, \lean{q}, and \lean{A.card}. We prove this via a generic result that holds for every \lean{zero_set_subspace} of a finite-dimensional space.

\begin{lstlisting}
theorem lemma_9_2 (T : subspace α (mv_polynomial (fin n) α))
  (A : finset (fin n → α)) :
  (vector_space.dim α zero_set_subspace).to_nat + A.card ≥
    (vector_space.dim α T).to_nat
\end{lstlisting}

This lemma is an exercise in linear algebra. It follows quickly from the rank-nullity theorem. The formal proof takes little work with our additions to the linear algebra theory in \mathlib.

We now set a parameter \lean{d : ℚ} which will remain fixed until the end of this section. After specializing \lean{lemma_9_2} and performing a cardinality computation, we obtain the following:
\begin{lstlisting}
theorem lemma_12_2 : q^n + V_dim ≥ m d + A.card
\end{lstlisting}

The \mathlib\ definition of \lean{vector_space.dim} takes values in the type \lean{cardinal}, since vector spaces are not restricted to finite dimensions. (Perhaps confusingly, \lean{finset.card} and \lean{fintype.card} take values in \lstinline{ℕ}.) In our setting, the vector space \lean{S'}, and hence its subspace \lean{V}, is finite dimensional. The cast \lean{cardinal.to_nat} is thus well behaved.

\subsection{Lemmas 2 and 3: Bounding the Dimension from Above}

Next we establish an upper bound for the dimension of \lean{V}. It is conceptually clearest to achieve this via two lemmas, one which bounds the dimension above by an intermediate value, and one which bounds this value above by \lean{m}.

To prove the first lemma, we define the support set of a polynomial to be the set of points on which it does not evaluate to 0:

\begin{lstlisting}
def sup (p : mv_polynomial (fin n) α) : finset (fin n → α) :=
finset.univ.filter (λ x, p.eval x ≠ 0)
\end{lstlisting}

A general argument about finite sets shows that there is some polynomial in \lean{V} with maximal support.
\begin{lstlisting}
lemma exi_max_sup :
  ∃ P ∈ V, ∀ P' ∈ V, sup P ⊆ sup P' → sup P = sup P'
\end{lstlisting}
We define \lean{P} to be this polynomial and \lean{P_sup} to be \lean{sup P}, allowing us to state the following:
\begin{lstlisting}
theorem lemma_12_3 : P_sup.card ≥ V_dim
\end{lstlisting}
The proof of this lemma involves some algebraic manipulation of the evaluation function \lean{mv_polynomial.eval}. It invokes yet another polynomial subspace, the zero set of \lean{V} with respect to \lean{P_sup}.

In order to relate \lean{P_sup} to other more interesting constants, we must prove a second lemma:

\begin{lstlisting}
theorem lemma_12_4 : P_sup.card ≤ 2 * m (d/2)
\end{lstlisting}
This lemma is a special case of Proposition \ref{prop 2 E-G} (Section \ref{section:background}), stated here in Lean:
\begin{lstlisting}
theorem proposition_11_1 {p :  mv_polynomial (fin n) α}
  (A : finset (fin n → α)) : p ∈ S' n d →
  (∀ (x : fin n → α), x ∈ A → ∀ (y : fin n → α), y ∈ A →
    x ≠ y → p.eval (a • x + b • y) = 0) →
  (A.filter (λ x, p.eval (-c • x) ≠ 0)).card ≤ 2 * m (d / 2)
\end{lstlisting}

Proving this proposition requires the most intricate argument of our formalization. We note that this is in line with Ellenberg and Gijswijt's paper; their corresponding Proposition~2 makes up nearly a third of the non-expository content. Some of the intricacy comes from another shift of representation. Every student of linear algebra learns that linear transformations between finite-dimensional vector spaces can be represented by matrices, and it is standard in mathematics to conflate the two concepts. While our lemma (after unfolding the definition of \lean{P_sup}) is stated in terms of the linear transformation \lean {p.eval}, Ellenberg and Gijswijt's argument proceeds more naturally in the matrix setting. Formalizing their argument required significant library development to unify the treatment of linear transformations and matrices in Lean. We expect that this development will be reusable in future results that depend on linear algebra.

Briefly, the proof of \lean{proposition_11_1} proceeds as follows. Given terms \lean{a b : α},\linebreak[4] \lean{x y : fin n → α}, and \lean{p : mv_polynomial (fin n) α} with \lean{p ∈ S' d}, the term \lean{p.eval (a • x + b • y)} can be written as a linear combination of evaluated monomials in \lean{M' d}. We define an \lean{A × A} matrix \lean{B} such that \lean{B x y = p.eval  (a • x + b • y)}. In fact, we can factor the matrix \lean{B} and express it in the following form:
\begin{lstlisting}
lemma B_eq_sum_matrix : B =
  split_left.sum (λ _ _, matrix.vec_mul_vec _ _) +
  split_right.sum (λ  _ _, matrix.vec_mul_vec _ _)
\end{lstlisting}
(We direct interested readers to our formalization for the details of this computation.) Here, the cardinalities of the finite sets \lean{split_left} and \lean{split_right} are at most \lean{m (d/2)}. Since the product of two vectors \lean{matrix.vec_mul_vec} has rank 1, this implies that \lean{B} has rank at most \lean{2 * m (d / 2)}. But in fact, \lean{B} is a diagonal matrix, from which we can infer that its rank is equal to the cardinality we wish to bound.

\subsection{Lemma 4: A Combinatorial Calculation}
\label{subsection:bound:combinatorics}

Our next lemma, largely independent of the previous ones, relates different values of \lean{m}.

\begin{lstlisting}
theorem lemma_12_5 : q^n ≤ m ((q-1)*n - d) + m d
\end{lstlisting}

This lemma follows from a combinatorial argument on \lean{fin n → fin q}, the type of \lean{n}-tuples of natural numbers less than \lean{q}. First, we define functions to map such a tuple to the monomial with corresponding exponents, and in reverse:
\begin{lstlisting}
def monom : (fin n → fin q) → mv_polynomial (fin n) α
def monom_exps : mv_polynomial (fin n) α → (fin n → fin q)
\end{lstlisting}
Note that these functions are inverses when we restrict \lean{fin n → fin q} to the subset \lean{M}.

We then define five terms of type \lean{finset (fin n → fin q)}, including the universal set:
\begin{itemize}
 \item \lean{I := finset.univ}
 \item \lean{B := \{v ∈ I // (total_degree (monom v)) ≤ d\}}
 \item \lean{C := \{v ∈ I // (total_degree (monom v)) > d\}}
 \item \lean{D := \{v ∈ I // (total_degree (monom v)) < (q-1)*n - d\}}
 \item \lean{E := \{v ∈ I // (total_degree (monom v)) ≤ (q-1)*n - d\}}
\end{itemize}

There are a number of straightforward cardinality calculations that follow. Among them, we show that \lean{B.card = m d}, since \lean{B} is the image of \lean{M' d} under \lean{monom_exps}. It similarly holds that \lean{E.card = m ((q-1)*n - d)}. The function sending the tuple $(a_1, \ldots, a_n)$ to $(q - 1 - a_1, \ldots, q - 1 - a_n)$ is a bijection and maps \lean{C} to \lean{D}; thus these sets have the same cardinality. Combining these calculations leads us to our goal.

Thanks to the large library of \lean{finset} operations in \mathlib, the proof of this lemma is basically frictionless. Indeed, the least pleasant part is checking that the bijection used is in fact a bijection, an argument that involves some trivial natural number arithmetic.

\subsection{Lemma 5: Connecting These Lemmas}

We have nearly achieved our goal for this section. Combining the previous four lemmas via linear arithmetic, we obtain the following:
\begin{lstlisting}
theorem lemma_12_6 : A.card ≤ 2 * m (d/2) + m ((q-1)*n - d) :=
by linarith using [lemma_12_2, lemma_12_3, lemma_12_4, lemma_12_5]
\end{lstlisting}
Finally, abstracting the parameter \lean{d} and instantiating it with \lean{2/3*(q-1)*n} delivers our desired bound.
\begin{lstlisting}
theorem theorem_12_1 : A.card ≤ 3*(m (1/3*((q-1)*n)))
\end{lstlisting}

\section{Asymptotics}
\label{section:asymptotics}

We have shown an upper bound for the cardinality of a cap set \lean{A} in terms of \lean{n}. To be precise, this bound is proportional to the number of monomials in \lean{n} variables with total degree at most \lean{(q-1)*n/3}, where \lean{q} is the cardinality of the underlying finite field.

Our goal was to investigate the growth rate of this bound, in terms of \lean{n}. In particular, we would like to show that it grows at a rate bounded above by \lean{c^n}, for some \lean{c < q}. Ellenberg and Gijswijt apply Cram\'er's theorem, a fairly deep result in probability theory (not to be confused with Cramer's rule), to derive this fact. But this detour is not necessary, and formalizing Cram\'er's theorem would be a significant undertaking on its own. We verify the growth rate of the size of \lean{A} using more elementary methods. While the results of this section could be stated in terms of $\mathcal{O}$-notation~\cite{affeldt:18}, we favor a more explicit style, which allows us to state the $q=3$ result in very concrete terms.

Our goal is the following general statement:
\begin{lstlisting}
theorem general_cap_set {α : Type} [discrete_field α] [fintype α] :
  ∃ B C : ℝ, B > 0 ∧ C > 0 ∧ C < card α ∧
    ∀ {a b c : α} {n : ℕ} {A : finset (fin n → α)},
      c ≠ 0 → a + b + c = 0 →
      (∀ x y z ∈ A, a • x + b • y + c • z = 0 → x = y ∧ x = z) →
      A.card ≤ B * C ^ n
\end{lstlisting}

Our motivating example is concerned with the case where the underlying field is $\ZZ/3\ZZ$. In this case, we can be more explicit about the growth rate: 
\begin{lstlisting}
theorem cap_set {n : ℕ} {A : finset (fin n → ℤ/3ℤ)} :
  (∀ x y z ∈ A, x + y + z = 0 → x = y ∧ x = z) →
    A.card ≤ 3 * (((3/8) ^ 3 * (207 + 33 * sqrt 33)) ^ (1/3)) ^ n
\end{lstlisting}
Since we have that
$$\sqrt[3]{\left(\frac{3}{8}\right)^3\left(207 + 33\sqrt{33}\right)}\, \approx\, 2.755,$$
this result answers the cap set problem in the affirmative. 

To prove \lean{general_cap_set}, we will show an alternate representation for \lean{m} and develop an argument that bounds this value from above in terms of \lean{n} and \lean{d}. This argument involves some combinatorial calculations similar to those presented in Section~\ref{subsection:bound:combinatorics}.

In the previous section we worked with a fixed parameter \lean{n}, the length of the vectors. It is now necessary to abstract over this parameter. (We will keep the base field \lstinline{α} and its cardinality \lean{q} fixed.) Note that \lean{m} depends on both \lean{n} and a rational input \lean{d}.

\subsection{Expressing \texttt{m} as a Sum of Coefficients}

Our first lemma will show that we can write \lean{m} as a sum of coefficients depending on $n$ and $d$. On paper, we define
$$ c_j^{(n)} := \left| \left\{ (a_1,\ldots,a_n) \enspace\middle\vert\enspace a_i \in \left\{0,1,\ldots,q-1\right\} \text{ and } \sum_{i=1}^n a_i = j \right\} \right|. $$

We again face a choice of how to represent these values in Lean. In Section~\ref{subsection:bound:combinatorics}, we represented such tuples $(a_1, \ldots, a_n)$ with the type \lean{fin n → fin q}. This type is very convenient when \lean{n} is fixed, but a following lemma will proceed by induction on \lean{n}, and the function representation is cumbersome in this kind of argument. We choose instead to represent these tuples with the type \lean{vector (fin q) n}, defined to be the subtype of \lean{list (fin q)} whose elements have fixed length \lean{n}. To connect with earlier results stated using the function representation, we will show a bijection between the two types. Moving between representations like this is aided by library support for establishing bijections and showing that relevant properties are preserved, and with the right support, it is far easier to carry out arguments in the ``natural'' setting.

With this in mind, we define:
\begin{lstlisting}
def sf (n j : ℕ) : finset (vector (fin q) n) :=
finset.univ.filter (λ f, (f.nat_sum = j))

def cf (n j : ℕ) : ℕ := (sf n j).card
\end{lstlisting}

Following the bijection between representations of tuples, and reusing some of the cardinality computations from Section~\ref{subsection:bound:combinatorics}, we show that \lean{m n d} is equal to the sum of \lean{cf q n j} for \lean{0 ≤ j ≤ ⌊d⌋}:

\begin{lstlisting}
theorem lemma_13_8 (n : ℕ) {d : ℚ} (hd : d ≥ 0) :
  m n d = (finset.range (⌊d⌋.nat_abs + 1)).sum (cf n)
\end{lstlisting}

To get a better handle on \lean{m}, we would like a more algebraic representation of \lean{cf}. As an intermediate step, we turn again to the setting of polynomials, this time univariate: we will show that for each $j$ and $n$, $c_j^{(n)}$ is equal to the $j$th coefficient of the polynomial $(1 + x + \ldots + x^{q-1})^n$.

It is in this argument that we benefit from using the list representation for tuples, as we need to prove:
\begin{lstlisting}
lemma cf_mul (n j : ℕ) : cf (n+2) j =
  (finset.range (j + 1)).sum (λ i, (cf 1 (j - i)) * cf (n + 1) i)
\end{lstlisting}
This combinatorial puzzle requires lifting $(n+1)$-tuples to $(n+2)$-tuples. Any $(n+2)$-tuple of natural numbers less than $q$ whose values sum to $j$ can be constructed by appending its last value $k$ to an $(n+1)$-tuple whose values sum to $i = j - k$. The number of such $(n+2)$-tuples, then, is the sum of the number of such $(n+1)$-tuples where $i$ ranges from 0 to $\max(q-1, j)$. Since \lean{cf 1 k} is 0 when \lean{k > q} and 1 otherwise, this sum is equal to the expression in \lean{cf_mul}.

Counting arguments like this can make for entertaining puzzles on paper, but the pain of formalizing them can be compounded by using the wrong representation. We found that the lifting of tuples required for this argument was much more natural under the list representation for tuples; casts in the function representation became unwieldy.

With this identity, and proceeding by induction on \lean{n}, we can define the polynomial $1 + x + \ldots + x^{q-1}$ and show our desired result:

\begin{lstlisting}
def one_coeff_poly (m : ℕ) : polynomial ℕ :=
(finset.range m).sum (λ k, (polynomial.X : polynomial ℕ) ^ k)

theorem lemma_13_9 (hq : q > 0) :
  ∀ n j : ℕ, ((one_coeff_poly q) ^ n).coeff j = cf n j
\end{lstlisting}

\subsection{Concrete Bounds on \texttt{m}}
\label{subsection:asymptotics:concrete}

We can now write \lean{m} in terms of the coefficients \lean{cf}. We will use this representation to establish a concrete upper bound on the values of \lean{m}. This upper bound will be in terms of another auxiliary value:
 \begin{lstlisting}
 def crq (r : ℝ) (q : ℕ) :=
 ((one_coeff_poly q).eval₂ coe r) / r ^ ((q-1)/3)
 \end{lstlisting}
Note that for \lean{p : polynomial ℕ} and \lean{r : ℝ}, \lean{p.eval₂ coe r} embeds the coefficients of \lean{p} into the real numbers and evaluates the resulting polynomial at \lean{r}. 

For every \lean{r} between 0 and 1, \lean{crq} bounds \lean{m}:
\begin{lstlisting}
theorem theorem_14_1 {r : ℝ} (hr : 0 < r) (hr2 : r < 1) : m ((q - 1)*n / 3) ≤ (crq r q) ^ n
\end{lstlisting}

This result is derived from \lean{theorem_13_8} and \lean{theorem_13_9}, with the additional fact that summing the monomials of a polynomial over its support is the same as evaluating the polynomial.
\begin{lstlisting}
lemma finset_sum_range {r : ℝ} (hr : 0 < r) (hr2 : r < 1) :
  (finset.range ((q - 1) * n + 1)).sum (λ j, r ^ j * (cf q n j)) =
    ((one_coeff_poly q) ^ n).eval₂ coe r
\end{lstlisting}

 Since \lean{crq 1 q = q} and the derivative of \lean{crq} with respect to \lean{r} is positive at \lean{r = 1}, we have from elementary calculus:
 \begin{lstlisting}
 theorem lemma_13_15 : ∃ r : ℝ, 0 < r ∧ r < 1 ∧ crq r q < q
 \end{lstlisting}
 Instantiating \lean{theorem_14_1} with this \lean{r}, invoking \lean{theorem_12_1}, and abstracting the type parameter \lstinline{α} leads us to the theorem \lean{general_cap_set} stated at the beginning of this section.

 We finally return to the original cap set problem with \lean{q = 3}. Pen and paper calculations show that \lean{crq r 1} is minimized in \lean{r} at \lean{r := (real.sqrt 33 - 1) / 8}. Aided by the numeral and ring normalization tactics in \mathlib, we establish that \lean{0 < r < 1} and that
 \lean{crq r 3 = ((3 / 8)^3 * (207 + 33*real.sqrt 33))^(1/3)}. We apply \lean{theorem_14_1} to this \lean{r} to conclude:
\begin{lstlisting}
theorem cap_set {n : ℕ} {A : finset (fin n → ℤ/3ℤ)} :
  (∀ x y z ∈ A, x + y + z = 0 → x = y ∧ x = z) →
    A.card ≤ 3 * (((3/8) ^ 3 * (207 + 33 * sqrt 33)) ^ (1/3)) ^ n
\end{lstlisting}

\section{Related Work}
\label{section:related}

We are not aware of any existing formal developments that relate directly to the cap set problem or the polynomial method. Since the core library components of our proof are in combinatorics and number theory, linear algebra, and the theory of polynomials, we provide here a survey of formalizations in these areas. This incomplete list is meant to indicate the depth and flavor of such projects.

The combinatorial arguments we employ are fairly simple results about involutions and the cardinalities of finite sets; similar developments exist in the libraries of most modern proof assistants. Gonthier's proof of the four color theorem in Coq~\cite{gonthier:08} includes some more sophisticated proofs. Dubois, Giorgetti, and Genestier~\cite{dubois:16} also provide a Coq library for enumerative combinatorics, again more sophisticated than what is needed in our proof.

While the result of Ellenberg and Gijswijt is most clearly characterized as combinatorics, it is also of interest in number theory. There has been recent attention toward formalizing results in this area, including Eberl's work on analytic number theory in Isabelle/HOL~\cite{eberl:19} and Lewis' work on the $p$-adic numbers in Lean~\cite{lewis:19}. Chyzak, Mahboubi, Sibut-Pinote, and Tassi's Coq proof that $\zeta(3)$ is irrational~\cite{chyzak:14} is also relevant.

Finite fields play an important role in combinatorics and number theory and are needed to state our general result. Chan and Norrish's mechanization of the AKS algorithm~\cite{chan:15} shows an approach to their study in HOL4, which makes for an interesting contrast with our approach in a dependently typed system. Their subsequent work~\cite{chan:16} relates to ours in its study of polynomials over finite fields. 

There are many formal proof developments of linear algebra. Our additions to \mathlib\ were partially inspired by the impressive work of Gonthier in Coq~\cite{gonthier:11}, Lee~\cite{lee:14} and Aransay and Divas\'on~\cite{aransay:14, divason:13} in Isabelle/HOL, and Harrison in HOL Light~\cite{harrison:13}.

Our formalization focuses in particular on the vector space of polynomials, also seen in Divas\'on, Joosten, Thiemann, and Yamada~\cite{divason:17}. 
As with linear algebra, polynomials are a fundamental object of study in mathematics, and they appear in most proof assistant libraries. Some recent results concerning polynomials include Bernard, Bertot, Rideau, and Strub~\cite{bernard:16} and Eberl~\cite{eberl:18}.

\section{Conclusion}
\label{section:conclusion}

We have formalized Ellenberg and Gijswijt's solution to the cap set problem, a recent and celebrated result in combinatorics. Our formalization is evidence that verifying certain cutting-edge mathematics is possible without enormous investments of time or resources. This effort was undertaken as part of the Lean Forward project, which aims to develop tools, tactics, and libraries to formalize modern results in number theory and related areas. Much of the background theory we have implemented will be of future use in this project.

At the outset of our efforts, the first author produced a detailed paper proof of the result, drawing from Ellenberg and Gijswijt and from Zeilberger~\cite{zeilberger:16} and adapting the asymptotics part significantly. We used this writeup as a blueprint for our formalization. It was heartening to see that the blueprint translated very directly to Lean. We were able to work at a similar level of abstraction as the original sources without any complications introduced by the proof assistant.

Our proof of the asymptotics is a significant simplification of the original arguments.
While in principle this could have been found without any interactive theorem proving, it was ultimately due to the formalization process, including the necessity to explore alternative paths of this part of the proof and feedback from Gijswijt on an earlier version of this paper, that this simplification was established.

As usual, it is difficult to compare the length of formal proofs with their paper counterparts, since the background assumptions and level of detail differ significantly. Nevertheless, we can provide some approximate information. Ellenberg and Gijswijt's paper contains just over two pages of mathematical work. Our blueprint is seventeen pages long; the first six pages are preliminary material, and two pages correspond to an obsolete argument (Appendix \ref{appendix}). The remaining nine pages correspond to around 2000 lines of our formalization. (This does not represent our entire effort: thousands more lines of general definitions and proofs were added to \mathlib\ as part of this project.) The ratio of 2000 lines of formal proof to two pages of paper proof is perhaps misleading, since we take a more verbose approach to checking the asymptotic behavior of the upper bound. (Ellenberg and Gijswijt take only one paragraph to invoke Cram\'er's theorem.) A better comparison is the part of the proof described in Section~\ref{section:bound}: 900 formal lines subsume a page and a half of paper proof. The corresponding section of our detailed writeup is just under five pages.

This formalization, and \mathlib\ more generally, rely heavily on hierarchies of type classes. In some sections of our proof---particularly those involving linear subspaces of the type of multivariate polynomials---we found that type class inference behaved erratically. The backtracking search performed by Lean's elaborator is sensitive to many features, and import order and additional instances can greatly affect the depth and speed of the search. We ended up revising the hierarchy in parts of \mathlib\ to simplify this. A moral we have taken from this project is that ``misleading'' instances that lead the elaborator down a long and ultimately unsuccessful path can be nearly as dangerous as circular instances.



\appendix

\section{An Earlier Proof of Asymptotics}\label{appendix}

After submission of our paper, Dion Gijswijt suggested a further simplification to the approach we used for controlling the asymptotic behavior of the bound. The argument we present above in Sections~\ref{section:background} and~\ref{section:asymptotics} follows this suggestion. For the sake of completeness, we present here our original approach, which may be of interest in its own right.

\subsection{Informal Description}

We will bound the coefficients of the polynomials from \eqref{eqn:comb id}:
\begin{equation}\label{eqn:comb id2}
m_d=\sum_{i=0}^{\lfloor d \rfloor} \left( \text{coefficient of $x^i$ in the polynomial } \left(1+x+\ldots x^{q-1}\right)^n \right).
\end{equation}
We can work in an algebraic manner as follows, thus avoiding Cauchy's residue theorem from complex analysis.
Let $k$ be any field, $f \in k[x]$, $i \in \NN$, $\zeta \in k^*$ of finite order $l$, and $r \in k^*$. If $l > \max (\deg(f),i)$, then
\begin{equation}\label{eqn:picking out coef}
l \cdot \left( \text{coefficient of $x^i$ in the polynomial $f$} \right) = \sum_{j=0}^{l-1} \frac{f(r \zeta^j)}{r^i \zeta^{ij}}.
\end{equation}
The key ingredient for proving this statement is the following special case of the geometric sum, where $\zeta$ and $l$ are as above and $h \in \ZZ$.
\[
\sum_{j=0}^{l-1} \zeta^{hj} =
\begin{cases}
0 & \text{if}\ l \nmid h \\
l & \text{if}\ l \mid h
\end{cases}
\]
Repeatedly applying~\eqref{eqn:picking out coef} to~\eqref{eqn:comb id2} with $k=\CC$, $\zeta=\exp(2 \pi \sqrt{-1} /l)$ for any $l > n(q-1)$, and $r \in \RR$ satisfying $0<r<1$, as well as calculating and estimating quite a bit, we obtain that
\[
m_{(q-1)n/3} \leq B_{r,q} C_{r,q}^n
\]
for some constants $B_{r,q}, C_{r,q} \in \RR_{>0}$ depending only on $r$ and $q$. Specifically, we can take
$C_{r,q}$ as in~\eqref{eqn:def Crq}.

\subsection{Formalization}

We pick up at the beginning of Section~\ref{subsection:asymptotics:concrete}, where we have not yet established an algebraic representation for \lean{cf}. It is necessary to get a better handle on the coefficients of \lean{one_coeff_poly ^ n}. A brief detour into estimates with complex numbers will result in the following bound:

\begin{lstlisting}
theorem lemma_13_10 (n : ℕ) {r : ℝ} (hr : r > 0) :
  cf n j ≤ (((one_coeff_poly q)^n).eval₂ coe r) / r^j
\end{lstlisting}
Note that for \lean{p : polynomial ℕ} and \lean{r : ℝ}, \lean{p.eval₂ coe r} embeds the coefficients of \lean{p} into the real numbers and evaluates the resulting polynomial at \lean{r}. This operation is generic, and we will soon embed this same polynomial into \lstinline{ℂ}.

To obtain the bound in \lean{lemma_13_10}, we will use a general result about complex polynomials. We derive this directly, but we note that it also follows from general considerations about Laurent polynomials:
\begin{lstlisting}
def ζk (k : ℤ) : ℂ := exp (2*π*I/k)

lemma pick_out_coef {f : polynomial ℂ} {i k : ℕ} (h1 : k > i)
  (h2 : k > nat_degree f) {r : ℝ} (h3 : r > 0) :
  (coeff f i) * k =
    (range k).sum (λ j, (eval (r*(ζk k)^j) f)/(r^i * (ζk k)^(i*j)))
\end{lstlisting}

When we instantiate \lean{f} with the embedding of \lean{one_coeff_poly ^ n} into \lstinline{ℂ}, we see that this complex sum is in fact a nonnegative real number for each \lean{i}, since it is equal to \lean{cf i n}. We can thus approximate its absolute value using the triangle inequality to derive \lean{lemma_13_10} above.

We can now write \lean{m} in terms of the coefficients \lean{cf}, and for each positive real \lean{r}, we can bound \lean{cf} from above in terms of \lean{r}. It remains to establish a concrete upper bound on \lean{m}.

We will do so using the same auxiliary value used in Section~\ref{subsection:asymptotics:concrete}:
\begin{lstlisting}
def crq (r : ℝ) (q : ℕ) :=
((one_coeff_poly q).eval₂ coe r) / r ^ ((q-1)/3)
\end{lstlisting}

It is convenient to first establish a bound in the case where \lean{n} is divisible by 3. The proof of this bound combines \lean{lemma_13_8} and \lean{lemma_13_10} with some elementary results about geometric sums.

\begin{lstlisting}
theorem lemma_13_11 (N : ℕ) {r : ℝ} (hr : 0 < r) (hr2 : r < 1) :
  m (3*N) ((q-1)*N) ≤ (1/(1-r)) * ((crq r q))^(3*N)
\end{lstlisting}

Recall that \lean{m n d} is the number of monomials in \lean{n} variables with total degree at most \lean{d}. This number is clearly monotonic increasing in \lean{d}; it is also easy to recognize that it is monotonic increasing in \lean{n}, although formalizing this takes slightly more work. From these considerations and the previous lemma, we deduce:
\begin{lstlisting}
theorem theorem_13_13 (n : ℕ) {r : ℝ} (hr : 0 < r) (hr2 : r < 1) :
  (m n ((q - 1)*n / 3)) ≤ ((crq r q)^2 / (1 - r)) * (crq r q)^n
\end{lstlisting}

As earlier, we can now derive from elementary calculus:
\begin{lstlisting}
theorem lemma_13_15 : ∃ r : ℝ, 0 < r ∧ r < 1 ∧ crq r q < q
\end{lstlisting}
Instantiating \lean{theorem_13_13} with this \lean{r}, invoking \lean{theorem_12_1}, and abstracting the type parameter \lstinline{α} leads us to the theorem \lean{general_cap_set}.

We finally return to the original cap set problem with \lean{q = 3}. Since we have used the same function \lean{crq} as in Section~\ref{subsection:asymptotics:concrete}, we can optimize it in \lean{r} in the same way to find the value \lean{r := (real.sqrt 33 - 1) / 8}. Aided by the numeral and ring normalization tactics in \mathlib, we establish that \lean{0 < r < 1} and that
\lean{crq r 3 = ((3 / 8)^3 * (207 + 33*real.sqrt 33))^(1/3)}. We compute the rough approximation \lean{(crq r q)^2 / (1 - r) ≤ 198} to conclude:

\begin{lstlisting}
theorem cap_set {n : ℕ} {A : finset (fin n → ℤ/3ℤ)} :
  (∀ x y z ∈ A, x + y + z = 0 → x = y ∧ x = z) →
    A.card ≤ 198 * (((3/8) ^ 3 * (207 + 33 * sqrt 33)) ^ (1/3)) ^ n
\end{lstlisting}

\bibliography{biblio}

\end{document}